# Films with the discrete nano-DLC-particles as the field emission cascade

Fengqi Song [a),*], Feng Zhou [b)], Haijun Bu [a)], Xiaoshu Wang [c)], Longbing He [b)],

Min Han [b)], Jianguo Wan [a)], Jianfeng Zhou [b)], Guanghou Wang [a)]

a) Department of Physics and National Laboratory of Solid State Microstructures, Nanjing University, 210093, P. R. China

b) Department of Material Science and Engineering and National Laboratory of Solid State Microstructures, Nanjing University, 210093, P. R. China

c) Center of Modern Analysis, Nanjing University, 210093, P. R. China

The films with the discrete diamond-like-carbon nanoparticles were prepared by the deposition of the carbon nanoparticle beam. Their morphologies were imaged by Scanning Electron Microscopy (SEM) and Atomic Force Microscopy (AFM). The nanoparticles were found distributed on the silicon (100) substrate discretely. The semisphere shapes of the nanoparticles were demonstrated by the AFM line profile. EELS was measured and the $sp^3$ ratio as high as 86% was found. The field-induced electron emission of the as-prepared cascade (nanoDLC/ Si) was tested and the current density of $1mA/cm^2$ was achieved at $10.2V/\mu m$.

* Corresponding author. Email: songfengqi@nju.edu.cn, ghwang@nju.edu.cn
  Telephone: +86-25 –83595082, Fax: +86-25-83595535

Field emission (FE) display was honored for its high resolution, small thickness and high color saturation, and was widely regarded as a strong contender of the LCD devices in the field of flat panel display. The structures and the field emission performances of the FE cascades are the unique factors in the total performance of the FEDs. The investigations on the FE mechanism and the cascade structure design have been started from the very early stage of the vacuum microelectronics [1, 2]. The energy level configuration by material optimization and the structure optimization by the morphology tuning have been both accepted as the two effective routes in FED engineering. In recent years, the engineering on working functions of the materials has transferred from metal-based materials to the Negative Electron Affinity (NEA) materials, such as diamond, carbon nanotubes (CNT) and cubic BN and diamond-like carbon (DLC) [3-6]. The configuration optimization has started a new era from the design of micrometer-sized Spindt-type [2] cascade to the structure design in nanoscale. The research on CNT FE has revealed that the structure of the nanosize tip of CNT played a more important role in the FE performance than the lengths [7, 8]. Therefore, the cascade with the same tip-shaped nanoparticles should be able to provide the comparable FE performance with the CNT-based ones as we suggest in this paper. Furthermore, the scanning deposition style of the nanoparticles may provide much larger emitter spatial density than the CNTs and more stable emission, for smaller heights of the emission sites and lower current density from a single emission site. Thus it may open the way to prepare the FE cascade by simple nanoparticle deposition and intermediate shape optimization. The DLC materials have been proved of NEA and excellent thermal stability and also mechanical properties [9]. In this letter the field emission of the asprepared nanoDLC/Si will be reported.

The schematic of the method was shown in Fig 1a. The samples were prepared by a cluster beam deposition method, which could provide the nanoparticle beams by a gas-aggregation process [10]. In the source the dense gas of carbon atoms came from the rf-magnetron sputtering and then composed nanoparticles by mutual collisions of the atoms. The base vacuum of the instrument was $2\times10^{-5}$Pa. The flux of the Argon flow was 92 sccm. The net input of the rf power was 300W. The beam formed by differential pumping with the pressure drop from 100Pa to $10^{-5}$Pa. The sizes of the clusters could be tuned from several atoms to several ten nanometers. The bonding composition and the structure of the nanoparticles can be engineered by changing the gas ambient and the beam-substrate interaction situations [10, 11], as schematically shown in Fig 1a. In this experiment, the beam of the nanoparticles with the diameters of 10-20nm was generated and then deposited onto the substrates including the TEM underlays and the silicon (100) sheets. The beam energy of 12keV was applied to enhance the cohesion between the nanoparticles and substrate with a group of planar electrodes. Fig 1b shows a SEM image of the asprepared sample. Some nanoparticles were found on the surface of n-type Si (100). They appeared round and the diameters were between 10-20nm. AFM was also carried out to reveal the 3d information. The line profile of a single nanoparticle was shown in Fig 1c. It has a diameter of about 21nm and a height of 11nm, smooth and symmetrical, nearly a shape of a semisphere.

The nanoparticles were also deposited onto some TEM underlays without carbon coating for TEM observation. The results of both electron diffraction and HR-TEM demonstrated the non-crystalline structure of the nanoparticles. Electron energy loss spectra (EELS) were measured to characterize the bonding situation within the nanoparticles. It confirmed the noncrystalline structure of the samples. The ratio of sp3 hybridization is very important in amorphous carbon materials, as it was known that a higher sp3 ratio will result in a more similar property to diamond, including NEA. The sp3 ratio can be obtained by the analysis of EELS result as shown in Fig 2. It is the core loss of carbon and the energy window is between 260~350eV. No other peaks appeared but the carbon band at 294eV, which proved the pure carbon composition of the sample. The ratio of $sp^3$ hybridization could be calculated from the spectrum by the method in ref [12]. A $sp^3$ ratio of 86% was achieved. It demonstrated the diamond-like bonding inside the nanoparticles.

As described above, our samples were of the structure with the mono-dispersed DLC nanoparticles (less than a monolayer). Its field emission performance was measured by a UHV FE instrument with a diode configuration, which has been described elsewhere [13]. Fig 3 showed the current density curve plotted against the electric field. The sample began its emission at the field of 4.6V/μm with a small current density of 2 nA/cm². The emission current grew fast with the increase of the applied electric field and reached 1mA/cm² at the field of 10.2V/μm. A current density of 2 mA/cm² could be reached at higher electric field. Such a total performance of the cascade is comparable to the CNT-based ones [14, 15]. The inserted red line in Fig 3 was the Fowler-Nordheim plot result of the I-V curve. A field-enhancement factor (β) of 400 was achieved. The emitted electrons were conducted to the phosphor screen coating by YOS:Eu with the red light as shown in the inset of the Fig 3.

It should be discussed that the straightness of the line indicated the Fowler-Nordheim emission mechanism of the cascade, based on field-induced electron tunneling through the interface potential. The value of the field enhancement factor (not equal to 1) revealed the existence of the local electric field enhancement due to the geometric structure of the emission sites (possible emitters). In the framework, the $sp^3$-hybridization-rich DLC nanoparticles/Si worked as the electron emitters, since the substrates are totally smooth except for the deposited nanoparticles. The contact between the nanoparticles and the substrate were strengthened because of the high beam energy, which weakened the obstacle between the nanoparticles and the substrate. And the tightly-contact $sp^3$-rich nanoparticles provided very low work function, which could be even zero for its possible NEA. That's why the electron emission occurred on the DLC-nanoparticle more easily than on the Silicon substrate. As to the present sample, the field enhancement factor of 400 fell highly behind the value of the CNT-based cascades. But the extremely high density of the possible emission sites ($10^{11}/cm^2$) retrieved the total performance of the cascade with a similar working level to the CNT-based one. Furthermore, the lower average current per emission site (of $10^{-14}$A per site) may result in less damage of the emitters, and essentially more stable working. It demonstrated the priority of the method of preparing the FE cascade by the nanoparticle deposition. Furthermore, the nanoparticle beam method realized the direct scanning deposition of the FE emitters, which provided the possibility to prepare large scale, high-emitter-density and uniform FE cascades [10]. It also provided the feasible acceleration and particle selection, which resulted in the improving of the nanoparticle cohesion and bonding optimization of the emission units respectively.

In conclusion, an FE cascade with the mono-dispersed sp3-hybridization-rich DLC-nanoparticles was prepared by energetic beam deposition and a good FE performance was demonstrated. The priorities of the method in the application of FE were emphasized.

## Acknowledgements

This work was financially supported by the National Natural Science Foundation of China (Grant Nos. 90606002, 10674056, 10674063, 10775070), NCET Project (NCET-07-0422), Hi-tech Research and Development Program of China under contract number 2006AA03Z316, the Foundation for the Author of National Excellent Doctoral Dissertation of P. R. China (Grant No. 200421). The authors are thankful for the help of Prof. Xiaoning Zhao of Nanjing Univ. and the Prof. Liping You of Peking Univ.

Figures' caption

1). A is the schematic of the method. The nanoparticles formed the energetic beams by the gas-aggregation source. And the structure of the nanoparticles could be tuned by changing the environment of the cluster growth. The scanning motion of the substrates resulted in the large scale of the scanning deposition. B is the Scanning Electron microscopy image of the sample, magnified by 300 000. The inset C is the AFM line profile result of a single particle. It has a semisphere shape. It is to be noted that the beam energy of 15keV was applied to enhance the cohesion between the nanoparticles and substrate.

2). The EELS of the sample. The ratio of sp3 hybridization is about 86%.

3). the result of the field emission measurement. The red inserted line was the F-N plot of the IV curve. The inset picture was the result of phosphors screen test by the powder of YOS: Eu. The total range of the picture is 0.2mm. It showed the situation at the very early stage of the field emission. The shadow came from the sample mask.

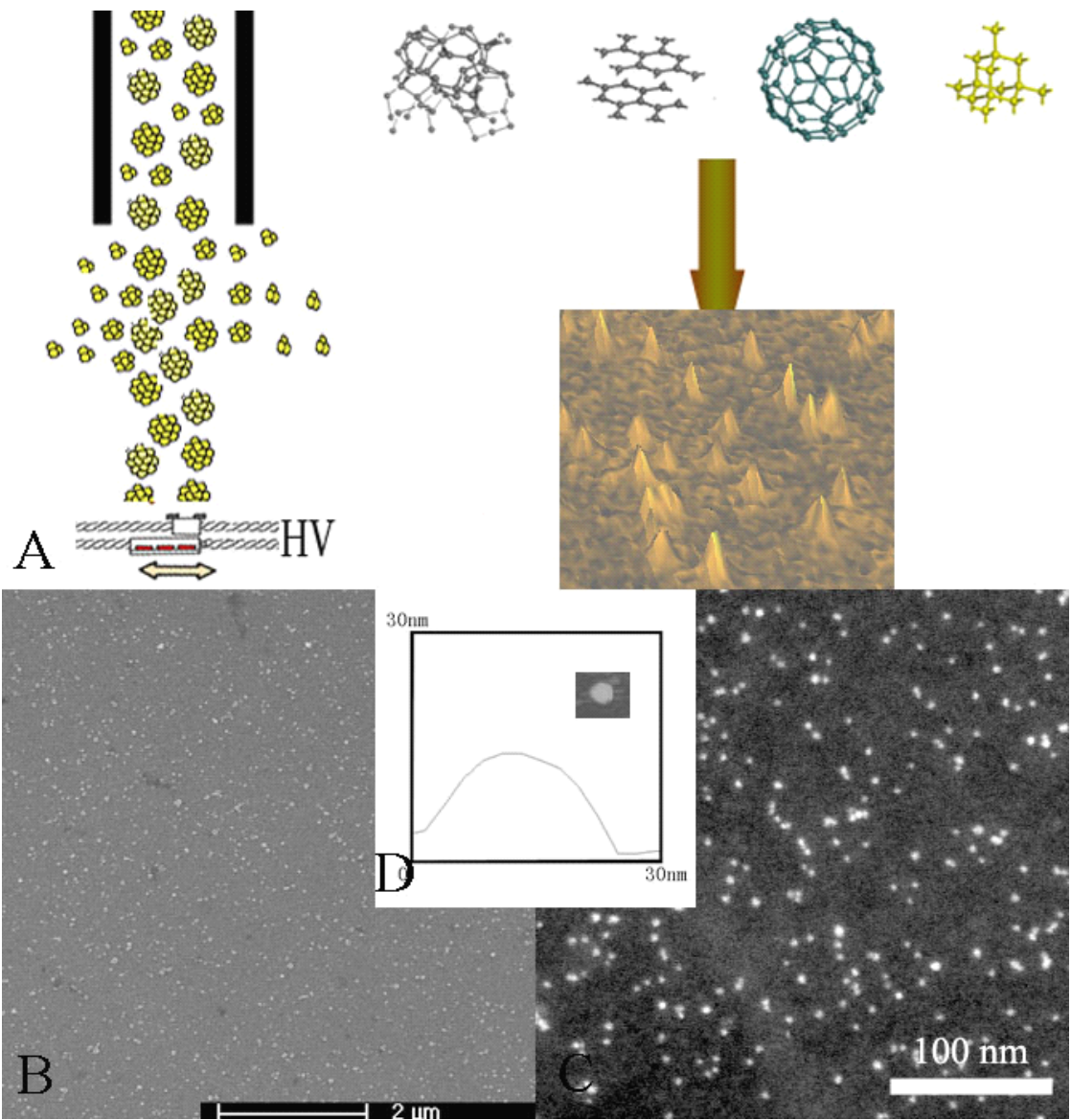

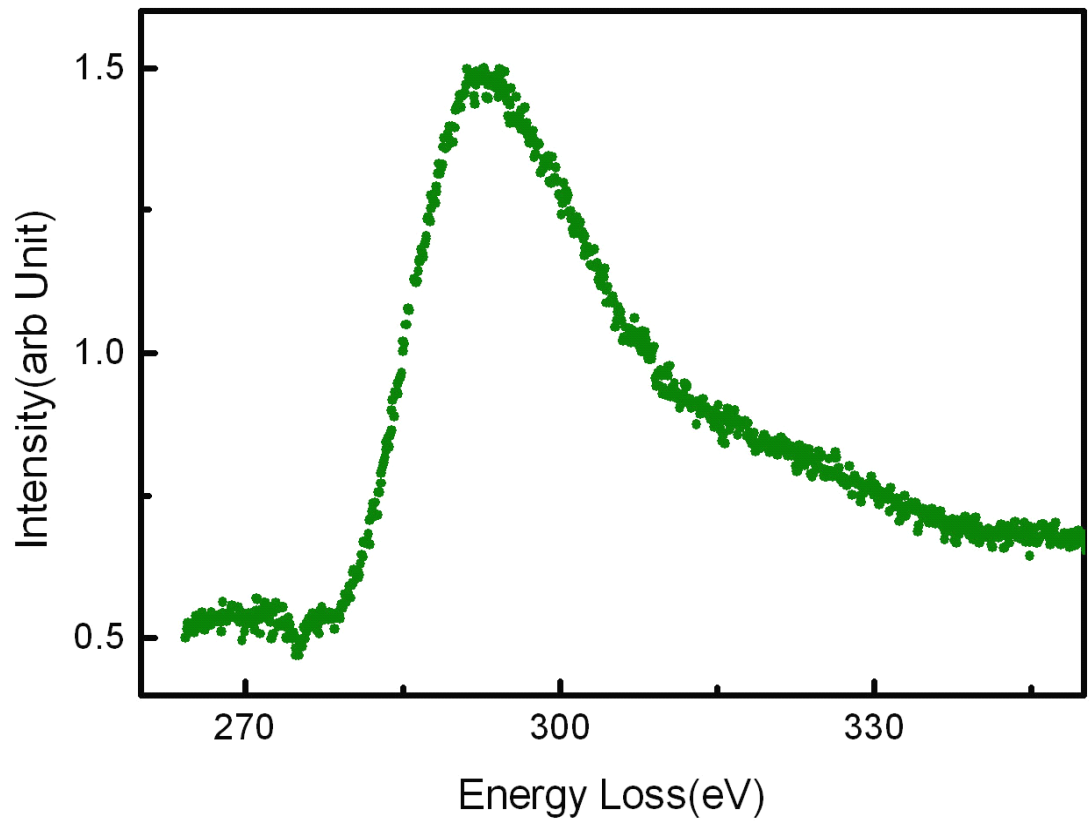

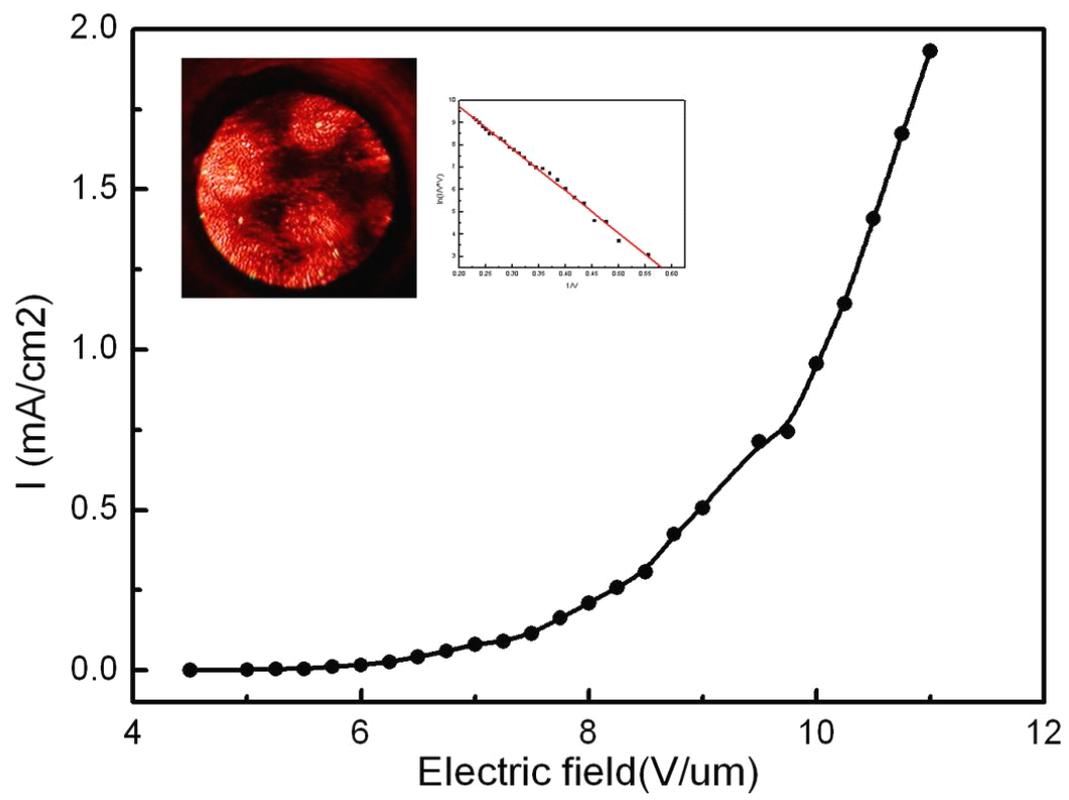